\documentclass[conference]{IEEEtran}
\IEEEoverridecommandlockouts
\usepackage{cite}
\usepackage{amsmath,amssymb,amsfonts}
\usepackage{algorithmic}
\usepackage{graphicx}
\usepackage{textcomp}
\usepackage{xcolor}
\usepackage{amsmath}
\usepackage{array}

\def\BibTeX{{\rm B\kern-.05em{\sc i\kern-.025em b}\kern-.08em
    T\kern-.1667em\lower.7ex\hbox{E}\kern-.125emX}}
\begin{document}

\title{News-Driven Stock Price Forecasting in
Indian Markets: A Comparative Study of Advanced
Deep Learning Models\\

}

\author{
\IEEEauthorblockN{Kaushal Attaluri\IEEEauthorrefmark{1}, Mukesh Kumar Tripathi\IEEEauthorrefmark{2}, Srinithi Reddy\IEEEauthorrefmark{3} and Shivendra\IEEEauthorrefmark{4}}
\IEEEauthorblockA{\IEEEauthorrefmark{1}\textit{Software Engineer-Machine Learning and AI}, \textit{Pegasystems India Worldwide Pvt Ltd}, Hyderabad, India\\ iamkaushal49@gmail.com}
\IEEEauthorblockA{\IEEEauthorrefmark{2}\textit{Department of Computer Science and Engineering}, \textit{Vardhaman College of Engineering}, Hyderabad, India\\ mukeshtripathi016@gmail.com}
\IEEEauthorblockA{\IEEEauthorrefmark{3}\textit{Department of Information Technology}, \textit{Vasavi College of Engineering}, Hyderabad, India\\ srinithireddy214@gmail.com}
\IEEEauthorblockA{\IEEEauthorrefmark{3}\textit{Department of Computer Application}, \textit{D. K. College}, Dumaraon, Bihar, India\\ srivastavashivendra29@gmail.com}
}

\maketitle

\begin{abstract}
Forecasting stock market prices is a challenging task for traders, analysts, and engineers due to the myriad of variables influencing stock prices. However, the advent of artificial intelligence (AI) and natural language processing (NLP) has significantly advanced stock market forecasting. AI's ability to analyze complex data sets allows for more informed predictions. Despite these advancements, stock price forecasting remains an area where AI has not yet achieved optimal results. In this paper, we forecast stock prices using 30 years of historical data from various national banks in India sourced from the National Stock Exchange. We employ advanced deep learning models, including multivariate multi-step Long Short-Term Memory (LSTM), Facebook Prophet with LightGBM and Optuna, and Seasonal Auto-Regressive Integrated Moving Average (SARIMA). Additionally, we analyze news data from tweets and reliable sources like Business Standard and Reuters, recognizing their significant impact on stock price movements.
\end{abstract}

\vspace{1\baselineskip}

\begin{IEEEkeywords}
Stock Market Forecasting, Data preprocessing, Advanced Algorithms, Deep Learning Models, Multivariate Multistep LSTM, SARIMA, LightGBM, Optuna, acebook Prophet,Sentiment Analysis.
    
\end{IEEEkeywords}

\section{Introduction}
It has always been challenging and highly competitive for traders and investors to optimize their earnings in the stock market by making sound judgments based on market trends and projections. Creating efficient prediction models is very important since these projections' precision significantly influences investment decisions. Various techniques, such as quantitative, fundamental, and technical analysis, have been developed to forecast stock prices[1-3]. Time series analysis is one of these that has become a potent tool for stock price prediction. Deep learning RNN architectures such as LSTM-based networks and conventional time series models like ARIMA have been used to forecast stock prices. Recently,  a new forecasting tool that has become very popular in data science is Facebook Prophet. Facebook created the Prophet time series forecasting model, which aims to produce precise and effective projections for various time series data, including stock prices [4-5].

The Indian stock market, like many other global financial markets, is susceptible to a variety of factors, including economic indicators, international events, and, increasingly, the vast amount of information disseminated through news media. In the digital age, news spreads rapidly, influencing investor sentiment and stock prices. The ability to accurately forecast stock prices based on news data has become a critical area of interest for researchers and market practitioners [6-7]. Through this comparative study, we aim to provide insights into the most effective deep-learning models for news-driven stock price forecasting in the Indian market. The findings will contribute to the growing body of knowledge in financial forecasting and could serve as a valuable tool for investors and financial analysts seeking to enhance their decision-making processes in a market heavily influenced by news and information flows.
This study aims to evaluate how well the enhanced Facebook Prophet, SARIMA, and Multivariate multistep LSTM models predict share value. The three models will be trained and assessed using historical stock price data from a selected company, with their performance measured by metrics such as Root Mean Squared Error (RMSE). In addition, we'll examine how these models stack up against LSTMs, which are attention-based and are trained not only on stock price data but also on news data from sources, including tweets and news articles from the Business Standard and Reuters. These methods will give us an idea of how the models work for Indian Stock Market Prices; this paper will also act as an understanding of Finance and Machine Learning
Enthusiasts. 
The proposed work has been discussed and will show the
several advantages, which will Section 3 of this paper,
but before that, Section 2 deals with the Related Work
that was carried out in the field of stock price prediction.
The other part of Section 3 will also deal with the
model’s architecture, process flow, and the rest of the
paper from Section 4 deals with the Results, Summary,
Future Work and References.
\section{related work}

In recent years, news-driven stock price prediction has drawn much attention as researchers try to use the abundance of data supplied by news stories to raise the accuracy of stock price forecasts as the judgments vary with the daily news. Several research papers have explored various deep learning models and techniques to address this difficult task and analyze the market accurately. This section offers a comparative analysis of prominent documents that have been presented at international conferences and have significantly advanced the field of news-driven stock price forecasting in Indian markets.

The influence of news on stock price movements has been a topic of extensive research, particularly as the availability of digital news sources has surged [8]. Traditional models, such as the Efficient Market Hypothesis (EMH), suggest that all available information, including news, is quickly reflected in stock prices, leaving little room for predictive modeling. However, the complexity and volume of news data have challenged this notion, paving the way for more sophisticated approaches like deep learning. Recent studies have increasingly focused on applying deep learning models to analyze the impact of news on stock prices [9-10]. Among these, the Bidirectional Long Short-Term Memory (Bi-LSTM) model has emerged as a powerful tool due to its ability to capture the temporal dependencies in sequential data and its proficiency in processing the context from past and future time steps. This bidirectional processing capability allows the model to understand better the sentiment and relevance of news articles about stock price movements.
 Several researchers have explored the effectiveness of Bi-LSTM models in stock price forecasting. For instance, some studies have demonstrated that Bi-LSTM models outperform traditional unidirectional LSTM models and other machine learning techniques when predicting the direction of stock price changes based on news data. This is attributed to the Bi-LSTM's ability to learn complex patterns and dependencies in the data, which simpler models often miss [11-12]. 

The literature [13-14]indicates that sentiment analysis in conjunction with Bi-LSTM models further enhances prediction accuracy. Sentiment analysis is a technique that involves determining the emotional tone behind a series of words used to understand how people feel about a particular topic or issue. By incorporating sentiment scores derived from news articles, these models can better gauge the market's reaction to news, leading to more accurate forecasts. The combination of Bi-LSTM with sentiment analysis has been shown to improve the model's performance, particularly in volatile markets where news plays a crucial role in driving price changes[15]. Overall, the literature highlights the significant potential of Bi-LSTM models in capturing the impact of news on stock prices. These models represent a significant advancement in financial forecasting, offering a more nuanced understanding of how information flows influence market trends. The ongoing development and refinement of these models, which includes improving the model's ability to handle larger and more diverse datasets, continue to contribute to the evolving landscape of stock price prediction, especially in markets susceptible to news, such as the Indian stock market.

In this paper [15-16], the authors focused on forecasting the stock price for the next day, specifically targeting the closing price of the S P 500 index. To achieve this, they employed Long Short-Term Memory (LSTM) models, a type of recurrent neural network (RNN) well-suited for time series prediction due to its ability to capture long-term dependencies in sequential data. A notable aspect of this study is using nine unique parameters as predictors, including fundamental market data, macroeconomic data, and technical indicators. These parameters were carefully selected to provide a comprehensive view of the factors influencing stock market behavior. The findings of this study revealed that the single-layer LSTM model outperformed the multi-layer LSTM model in terms of fit and prediction accuracy [17]. Specifically, the single-layer model achieved lower RMSE and MAPE values, indicating a closer match between predicted and actual stock prices. The higher correlation coefficient in the single-layer LSTM model also suggested a stronger relationship between the expected and actual values.

Much of the paper is dedicated to an in-depth discussion of LSTM networks, which are particularly well-suited for time series forecasting due to their ability to capture long-term dependencies in sequential data. The authors explained the architecture and functioning of LSTM networks, highlighting their advantages over traditional models in handling the complexities of financial data. The authors [18] employed other machine learning algorithms, such as Random Forests, deep neural networks, and logistic regression. By integrating these diverse models, they developed a comprehensive trading approach that leverages the strengths of each method. The feature space and target variables for both training and prediction were identified. This step involved selecting relevant financial indicators and metrics that could serve as inputs to the model, enabling it to learn patterns and trends influencing stock prices.

A unique aspect of this study is employing a CNN model, where the authors utilized image data representing stock price trends. By converting time series data into images, the CNN model could capture latent dynamics and patterns within the data that might not be easily discernible through traditional numerical analysis. This innovative approach allowed CNN to extract features from the visual representation of stock price trends, providing a new dimension to financial data analysis. The study also incorporated a sliding window algorithm, a technique used to predict short-term future values [19]. This method uses a fixed-size window of past data to predict the next value in the sequence. The sliding window approach is efficient for capturing short-term trends and making near-future predictions, which are crucial for traders and investors looking for timely insights. The findings underscore the potential of deep learning models, particularly when innovative methodologies such as the sliding window algorithm and image-based analysis are integrated into the prediction process.

\section{Proposed Work}
Our research suggests a thorough and complex method for stock price prediction in Indian markets, taking the data from the news and combining several deep learning models, conventional time series analysis approaches, and natural language processing methods. The combination of several deep learning models, such as Prophet, Multistep LSTM, and SARIMA, forms the foundation of our idea. These architectures effectively identify the complex relationships and nonlinear patterns present in the Indian stock market data, improving our predictions' precision and robustness. 

To improve our model's accuracy even more, we provide a text segmentation module that uses the Hidden Markov Model (HMM) to identify and extract the most noteworthy and informative sections from news stories about the Indian stock market. By utilizing the Viterbi algorithm, we achieve accurate segmentation by concentrating only on crucial textual data that substantially influences changes in stock prices. In addition, we incorporate a Bidirectional LSTM-based text emotion classification module to fully assess news item emotional context, allowing for a more sophisticated comprehension of market mood and its possible impact on stock prices. Our model uses the rich contextual data that Bidirectional LSTMs gather to classify news articles into several emotion classes ( positive, negative, and neutral), increasing our method's potential for prediction. We aim to empirically validate the effectiveness and superiority of our proposed approach, highlighting its capacity to provide precise and valuable stock price predictions amidst the complexities and dynamics inherent in the Indian market landscape through rigorous experimentation and careful evaluation of real-world data from the Indian stock market. 
\begin{figure}[htbp]
    \centerline{\includegraphics[width=0.61\textwidth]{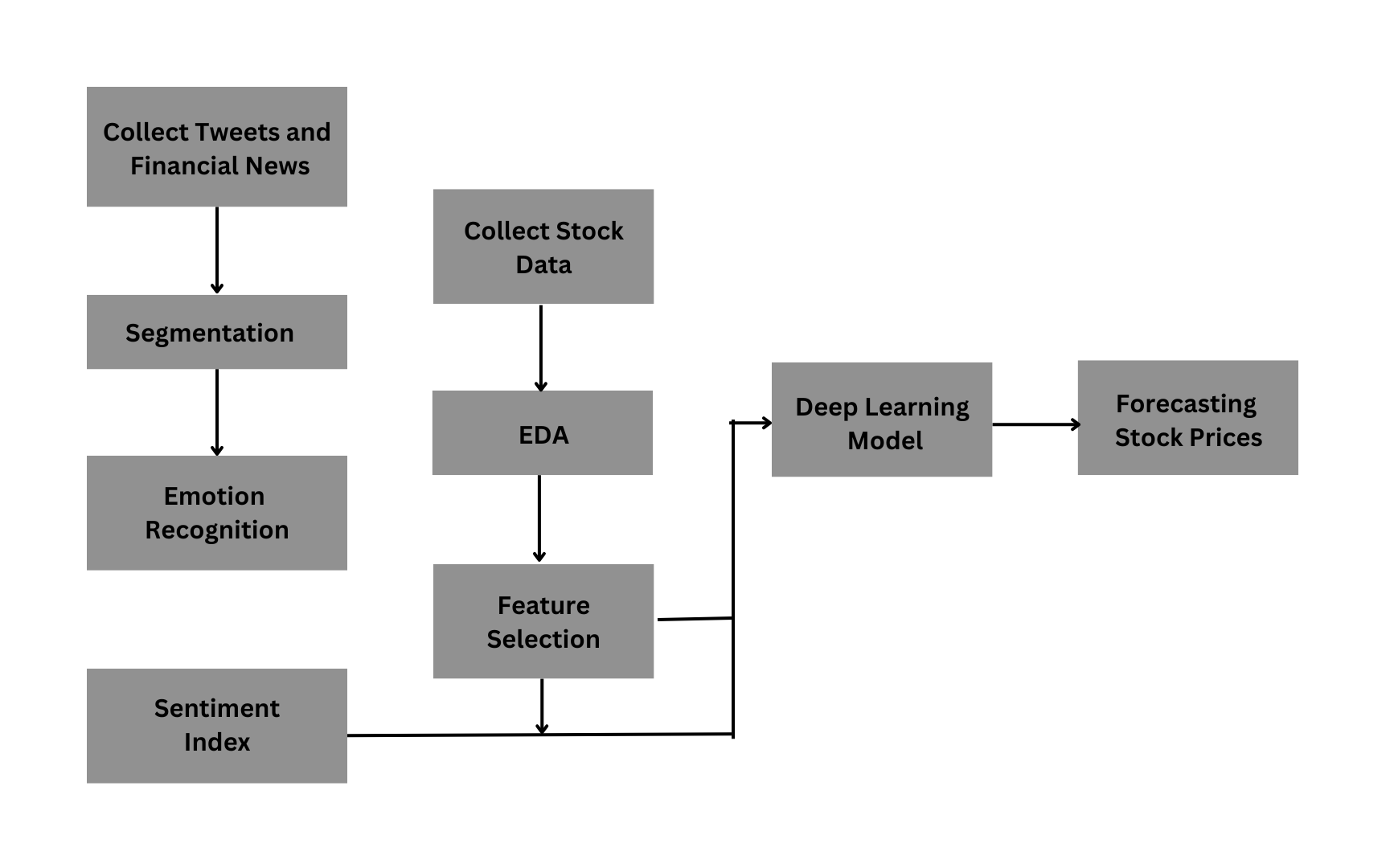}} 
    \caption{Block diagram}
    \label{fig}
\end{figure}
\section{Methodology}
\subsection{Data Preprocessing}
The Dataset used in this study will take the parameters
OLHCV (Open High Low Close Volume) prices of the
Stock. All the values have been taken from Yahoo Finance. We have chosen 4 different banks, 2 public sector and 2 private banks.
The period taken for this share price was from the last 20 years. 

We were initially using Pandas datareader and quandl APIs to get our data, but there were some issues with it while parsing so we have used yfinance which will easily fetch us these banks' National Stock Exchange prices.

\subsection{Multivariate Multistep LSTM MODEL}
Long Short-Term Memory (LSTM) is an advanced variant of the recurrent neural network (RNN) introduced by Hochreiter and Schmidhuber. In general, Time-series forecasting. we are predicting future dependent variables (y) based on the previous independent variables(x).
In univariate forecasting, y is predicted on that independent variable (x). In multivariate, there are many independent variables (x1,x2,x3,...xn), and here we employ multistep forecast, which means predicting a few times-steps.
So in our research, we employ four independent variables (x1,x2,x3,x4).

We are rearranging our data in split sequences, with two arguments: how much data we have to look back for our prediction and how much multistep data we want to forecast. (Let us denote them by input and output).
We are marking our input and output to be 8 and 9.
Here, we look back at the last 30 (x1,x2,x3,x4) and forecast the future for 30 multi-steps ahead. So, in total, we have 72 batches of dependent variables and independent variables. We are splitting them into 69 for training and 3 for testing.
The shapes of our training data and testing data should be (104820, 30, 4) and (16480, 30, 3). The author has to start our LSTM architecture by initializing the optimizer learning rate and number of layers. For this research, we have decided to go with the Adam optimizer.
\begin{figure}[htbp]
    \centerline{\includegraphics[width=0.5\textwidth]{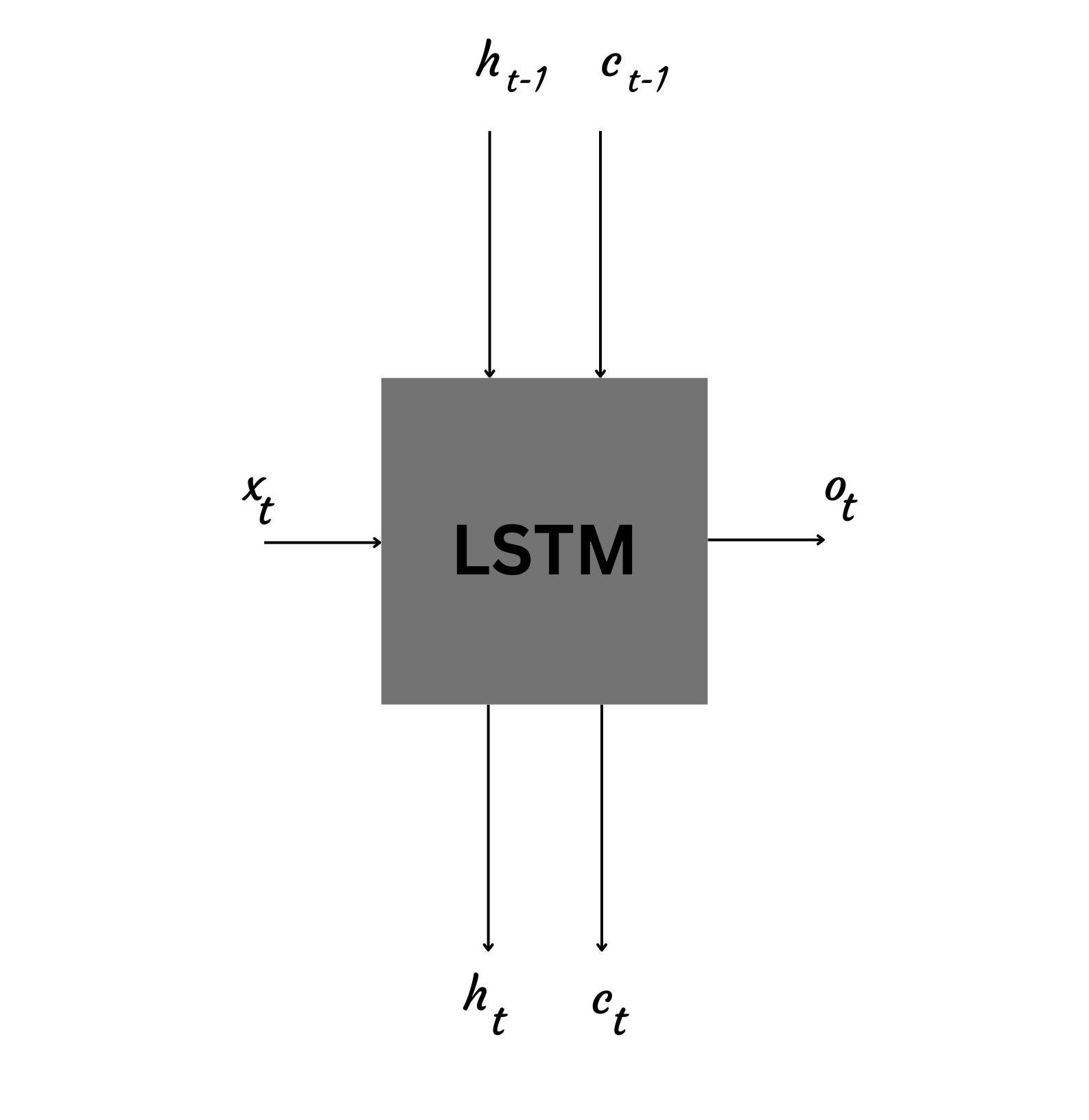}} 
    \caption{LSTM Basic Architecture}
    \label{fig}
\end{figure}
\\ 

\[
f_t = \sigma(W_f \cdot [h_{t-1}, x_t] + b_f)
\]
- \textbf{\( f_t \)}: The forget gate has an output entry, which determines which portions of the cell state should be discarded. \\
- \textbf{\( W_f \)}: Weight matrix for the forget gate operation. \\
- \textbf{\( [h_{t-1}, x_t] \)}: Combined vector of the prior hidden state and the current input. \\
- \textbf{\( b_f \)}: Bias term associated with the forget gate. \\

\[
i_t = \sigma(W_i \cdot [h_{t-1}, x_t] + b_i)
\]
- \textbf{\( i_t \)}: The input gate has an output, which decides which new information is incorporated into the cell state. \\
- \textbf{\( W_i \)}: Weight matrix for the input gate operation. \\
- \textbf{\( [h_{t-1}, x_t] \)}: Combined vector of the prior hidden state and the current input. \\
- \textbf{\( b_i \)}: Bias term associated with the input gate. \\

\[
\tilde{C}_t = \tanh(W_C \cdot [h_{t-1}, x_t] + b_C)
\]
- \textbf{\( \tilde{C}_t \)}: Candidate value for the state cell, representing potential updates. \\
- \textbf{\( W_C \)}: Weight matrix used for generating the candidate cell state. \\
- \textbf{\( [h_{t-1}, x_t] \)}: Combined vector of the prior hidden state and the current input. \\
- \textbf{\( b_C \)}: Bias term associated with the candidate cell state. \\

\[
C_t = f_t \odot C_{t-1} + i_t \odot \tilde{C}_t
\]
- \textbf{\( C_t \)}: Updated cell state, combining retained information from the previous state with new inputs. \\
- \textbf{\( f_t \)}: forget gate's output. \\
- \textbf{\( \odot \)}: Element-wise multiplication operator. \\
- \textbf{\( C_{t-1} \)}: the previous time step's cell state. \\
- \textbf{\( i_t \)}: Output of the input gate. \\
- \textbf{\( \tilde{C}_t \)}: Candidate cell state value. \\

\[
o_t = \sigma(W_o \cdot [h_{t-1}, x_t] + b_o)
\]
- \textbf{\( o_t \)}: The output gate's output controls the information output from the cell state. \\
- \textbf{\( W_o \)}: The output gate's weight matrix. \\
- \textbf{\( [h_{t-1}, x_t] \)}: Combined vector of the prior hidden state and the current input. \\
- \textbf{\( b_o \)}: Bias term for the output gate. \\

The role of the forget gate tells how much data for the previous cell state should be preserved for the current state cell. It picks the sigmoid activation function by default, ensuring the output is constrained to 0 and 1 [21]. The role of the input gate is to check what elements or data for the new input should be assigned for the cell state. Here, the activation function is also sigmoid by default, where the output range is between 0 and 1, which will signify the data that is coming for updating the cell state.The role of the candidate cell is to introduce all the new data that may or may not be added to the state cell. Here, we use a tanh activation function ranging between -1 and 1. The role of the output gate is to finally produce the output for each time step, which becomes the hidden state for the following time step. Here, the sigmoid will control all the output values from 0 to 1 and indicate the portion of the state cell's data available for the output.\\

\subsection{Facebook Prophet}
Prophet has been a widely used forecasting tool in time series developed by the Meta AI team . It is designed to handle a lot of data and assist businesses in understanding and predicting market trends. This open-source tool employs a decomposable additive model that accommodates non-linear trends, seasonality, and the impact of holidays. Before diving into practical applications, it’s essential to familiarize yourself with some key concepts:

\begin{itemize}

\item \textbf{Trend}: This is the main component of the Prophet architecture; this will represent the path of the underlying information over a while, effectively filtering out short-term fluctuations and seasonal effects.

\item \textbf{Seasonality}: Seasonality refers to repetitive variations over shorter intervals. These variations are less persistent than trends and are characterized by regular patterns within specific periods.

\end{itemize}

\[ 
y(t) = g(t) + s(t) + h(t) + \epsilon_t
\]

\[
g(t) = (k + a(t)^T \delta)t + (m + a(t)^T \gamma)
\]
\[
g(t) = \frac{C}{1 + \exp(-(k + a(t)^T \delta)(t - (m + a(t)^T \gamma)))}
\]

\[
s(t) = \sum_{k=1}^N \left[ a_k \cos\left( \frac{2 \pi kt}{P} \right) + b_k \sin\left( \frac{2 \pi kt}{P} \right) \right]
\]
- \textbf{\( N \)}: Seasonal components. \\
- \textbf{\( a_k, b_k \)}: Fourier coefficients. \\
- \textbf{\( P \)}: Period. \\
\[
h(t) = \sum_{j=1}^L \left[ \beta_j D_j(t) \right]
\]
- \textbf{\( L \)}: Number of holidays. \\
- \textbf{\( \beta_j \)}: Holiday effects. \\
- \textbf{\( D_j(t) \)}: Holiday indicators. \\
The author also combines the prophet model with the LightGBM, an open-source boosting framework that Microsoft has developed.
This is based on decision trees, which will improve the model's efficiency and accuracy. In this method, we are using LightGBM as a wrapper integrated with the Optuna Library, a hyperparameter optimization framework. Here, our overall method is a stepwise approach where we optimize the hyperparameters at a single time and employ a pre-defined search space. This approach is much faster than Grid search, where a combination of all parameters training takes place at once. 
In gridwisee the parameters will be h1*h2*h3*h4...hn but in our approach we'll have the total number of parameters to be h1+h2+h3+...hn, where hi value represents values for that parameter, Overall this approach will reduce the number of parameters and has increased our efficiency and accuracy.

\subsection{SARIMA}
The author also employs a seasonal autoregressive integrated moving average (SARIMA), an extended version of ARIMA and a popular time series forecasting tool. It analyzes the relationship between a dependent variable and its past values to predict future trends. Unlike methods that use actual values, ARIMA models focus on the differences between successive observations to forecast future financial or market movements.
SARIMA is denoted as ARIMA(a,b,c)(A, B, C)s. 
Here, the hyperparameters are denoted as 
a: This parameter indicates the Trend autoregression order
b: This parameter indicates the Trend difference in order
c: This parameter indicates the average order of the moving Trend

Some seasonal elements are generally not configured in ARIMA but must be configured for our model
A:  Order of Seasonal autoregression
B:  Order of Seasonal difference
C:  Order of Seasonal moving average
m:  Time steps for a single seasonal interval

\subsection{Sentiment Recognition}
In addition to traditional time-series forecasting models, sentiment recognition is crucial in enhancing the prediction accuracy of Indian bank stock prices. The sentiment data were meticulously gathered from various sources, including prominent news outlets like \textit{Business Standard} and \textit{Reuters}, as well as real-time Twitter tweets, ensuring comprehensive coverage of market sentiment. The author used the Hidden Markov Model (HMM) to analyze these sentiment data. The HMM framework allowed us to model the latent sentiment states, denoted by the set $\mathcal{L} = \{l_1, l_2, \dots, l_n\}$, which could not be observed directly but influenced the observable data sequences, represented by $\mathcal{ds} = \{ds_1, ds_2, \dots, ds_T\}$. The model parameters included the transition probability matrix $\mathbf{Ax}$, which governs the probability of moving from one sentiment state to another, and the observation probability matrix $\mathbf{Bx}$, which defines the likelihood of observing a particular sentiment indicator given a specific state. The vector $\boldsymbol{\pi}$ denoted the initial state distribution. We utilized the Viterbi algorithm to decode the most probable sequence of hidden sentiment states that could have generated the observed data. This algorithm efficiently computes the most likely path of sentiment states, $\mathcal{L}^* = \{l_1^*, l_2^*, \dots, l_T^*\}$, by maximizing the posterior probability $P(\mathcal{L}|\mathcal{ds})$ over all possible state sequences. This path reveals the temporal evolution of sentiment in the market, which is a critical factor in influencing stock price movements. The resulting sentiment scores, derived from the Viterbi-decoded state sequence, were then incorporated as exogenous variables into the stock price forecasting models. This integration allows the models to account for market sentiment, providing a richer and more informed prediction of future stock prices. By combining sentiment analysis with multivariate time-series models, we aim to capture both the quantitative and qualitative aspects of market dynamics, leading to more robust and accurate forecasting outcomes.

\section{RESULTS}\label{SCM}
\subsection{Results using LSTM}
The study calculated results for the LSTM model using the stock's closing prices, focusing initially on a univariate model. This univariate approach relied solely on the closing prices as the input data, providing a straightforward yet effective method for predicting future stock prices. The simplicity of this model makes it easy to implement, though it may need to capture the broader market dynamics that influence price movements. The authors also considered a multivariate model to enhance the predictive power, where a multistep multivariate LSTM was employed to predict stock prices over five future time steps. By incorporating these extra features, the model was able to capture more complex patterns and interactions within the stock market data. Including multiple variables in the multivariate model provided a more holistic view of the market conditions. The opening price and stock volume, in particular, are crucial indicators of market sentiment and trading activity, and their inclusion allowed the LSTM model to understand better and predict the factors driving stock price movements. This comprehensive approach aimed to improve the prediction accuracy compared to the univariate model by leveraging the interdependencies among different market indicators. 

\begin{figure}[htbp]
    \centerline{\includegraphics[width=0.5\textwidth]{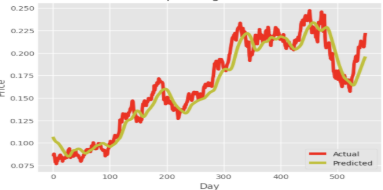}} 
      \caption{: Split 1 Multistep LSTM}
    \label{fig}
\end{figure}

 \begin{figure}[htbp]
    \centerline{\includegraphics[width=0.5\textwidth]{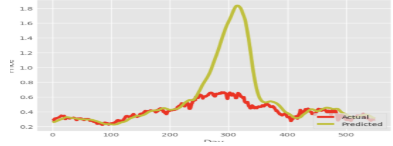}} 
      \caption{: Split 2 Multistep LSTM}
    \label{fig}
\end{figure}

\begin{figure}[htbp]
    \centerline{\includegraphics[width=0.5\textwidth]{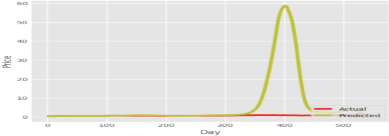}} 
      \caption{: Split 3 Multistep LSTM}
    \label{fig}
\end{figure}

\begin{figure}[htbp]
    \centerline{\includegraphics[width=0.5\textwidth]{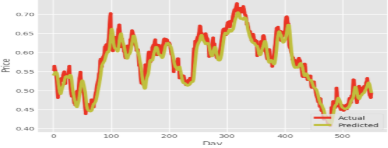}} 
      \caption{: Split 4 Multistep LSTM}
    \label{fig}
 \end{figure}
 
\begin{figure}[htbp]
    \centerline{\includegraphics[width=0.5\textwidth]{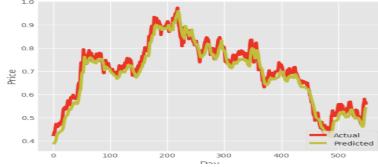}} 
      \caption{: Split 5 Multistep LSTM}
    \label{fig}
\end{figure}

\subsection{Results using Facebook Prophet}
In the study, the same period used for the LSTM model was also considered for evaluating the performance of the Prophet model, a popular tool for time series forecasting. The visualization of the results included the following elements.
Black Dots Represent the actual data points of the stock prices. These dots provide a reference for the exact values against which the model's predictions can be compared. The blue line shows the central tendency of the projections made by the Prophet model over the selected time, providing a clear indication of the predicted trend. Sky Blue Area Illustrates the variability in the model's predictions, often called the uncertainty interval or confidence interval.

\begin{figure}[htbp]
\centerline{\includegraphics[width=0.5\textwidth]{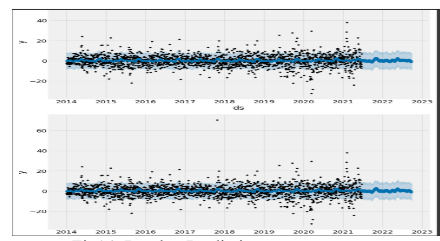}} 
\caption{Prophet Prediction}
\label{fig}
\end{figure}

\subsection{Results using SARIMA Model}
Here, the nonseasonal order is set as (1,1,1), and the seasonal order is set as (3,3,1,15). One thing we have observed with SARIMA is that it can predict stock prices well, even during seasonal trends. During COVID-19, around April-September 2020, SARIMA came up with the best forecasting as it quickly adjusted to the seasonal trend compared to other models.

\begin{figure}[htbp]
\centerline{\includegraphics[width=0.5\textwidth]{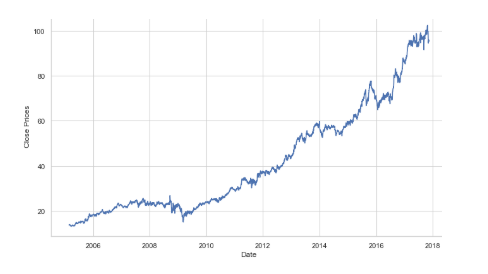}} 
\caption{Prophet Prediction}
\label{fig}
\end{figure}
\subsection{Evaluation Metrics}
\paragraph{Root Mean Square Error (RMSE)} Root Mean Square Error (RMSE) is a widely utilized metric for assessing the accuracy of predictions. It quantifies the deviation between predicted and actual observed values by calculating the Euclidean distance between them. To determine RMSE, follow these steps: first, compute the residuals for each data point, which is the difference between the predicted value and the actual value. Next, calculate the square of each residual, then find the average of these squared residuals. Finally, take the square root of this average. RMSE is particularly useful in supervised learning contexts as it requires valid values for each prediction.


The following table presents the RMSE values for four stock prices sourced from the National Stock Exchange, evaluated using various models.

\begin{table}[h!]
\caption{Model performance on different stocks}
\centering
\begin{tabular}{|>{\centering\arraybackslash}m{3cm}|c|c|c|c|}
\hline
\textbf{MODEL} & \textbf{Stock 1} & \textbf{Stock 2} & \textbf{Stock 3} & \textbf{Stock 4} \\
\hline
Univariate LSTM & 4.89 & 3.14 & 8.12 & 2.34 \\
\hline
Multivariate Multi-step LSTM (Global Average) & 3.91 & 2.990 & 5.96 & 1.98 \\
\hline
SARIMA & 11.28 & 10.281 & 14.37 & 9.87 \\
\hline
Facebook Prophet + LGBM + Optuna & 6.47 & 7.252 & 6.98 & 5.90 \\
\hline

\end{tabular}

\label{table:1}
\end{table}

\section*{Conclusion}
In conclusion, the study provides valuable insights into the effectiveness of various models for stock price forecasting, highlighting each approach's relative strengths and weaknesses. The attention-based Multistep LSTM, which integrates news sources and Twitter data, emerged as the top performer, demonstrating superior accuracy compared to other models. This suggests incorporating real-time sentiment and information from diverse sources can significantly enhance predictive performance. However, it is also noteworthy that the Multivariate LSTM model, which considers multiple financial indicators, performed consistently well across all cases, making it a reliable alternative when real-time data sources are unavailable. The findings indicate that each model has its strengths. They also emphasize the importance of continuous improvement and refinement in stock price forecasting methodologies. The results contribute to a broader understanding of how different models perform in real-world scenarios, offering valuable guidance for future research and practical applications in financial forecasting.

\vspace{12pt}

\end{document}